\documentclass[pre,superscriptaddress,twocolumn]{revtex4-1}

\usepackage{amsmath}
\usepackage{epsfig}
\usepackage{subfigure,mathrsfs}
\usepackage{array}

\newcommand{\beq}[0]{\begin{equation}}
\newcommand{\eeq}[0]{\end{equation}}
\def\be{\begin{equation}}
\def\ee{\end{equation}}
\def\bea{\begin{eqnarray}}
\def\eea{\end{eqnarray}}
 
\begin{document}

\title{Minimal universal quantum heat machine }

\author{D. Gelbwaser-Klimovsky}
\affiliation{Weizmann Institute of Science, 76100
Rehovot, Israel}
\author{R. Alicki}
\affiliation{Weizmann Institute of Science, 76100
Rehovot, Israel}
\affiliation{Institute of Theoretical physics and Astrophysics,
University of Gda\'nsk}
\author{G. Kurizki}
\affiliation{Weizmann Institute of Science, 76100
Rehovot, Israel}

\begin{abstract}
In traditional thermodynamics the Carnot cycle  yields  the ideal performance bound of heat engines and refrigerators. We propose and analyze a minimal  model of a heat machine that can play a similar role in  quantum regimes. The  minimal model consists of a single two-level system with periodically modulated energy splitting that is permanently,  weakly, coupled to two spectrally-separated heat baths at different temperatures. The equation of motion allows to compute the stationary power and heat currents in the machine  consistently with the second-law of thermodynamics. This dual-purpose  machine  can act as either  an engine or a refrigerator (heat pump) depending on the modulation rate. In both modes of operation the maximal Carnot efficiency is reached at zero power. We study the conditions for finite-time optimal performance for several variants of the model. Possible realizations  of the model are discussed.
\end{abstract}
 
\maketitle

\section{Introduction}
Emerging technologies aim at operation based on quantum mechanics \cite{RablPRL06,*ImamogluPRL09,*YouPT05,*MakhlinRMP01,*TaylorNAT08,*HenschelPRA10,*KaschNJP10,*FrancheschiNAT10,*TordrupPRA08,*DuttSCI07,*SillanpaaNAT07,*LeistikowPRL11,BenskyQIP11,*PetrosyanPRA09}, but  power-supply or cooling devices are still  governed by traditional (19th century) thermodynamics \cite{CallenBOOK85,*ReichlBOOK98}. It is therefore imperative to examine the conceptual compatibility of the two disciplines as regards the performance of such devices \cite{GeusicPR67,*GevaJMO02,*QuanPRE07,*VandenBroeckPRL05,*LinPRE03,*BenderJPA00,*EspositoPRL09,*QuanPRE07,*FialkoPRL12,AlickiOSID04,*AlickiJPA79,SegalPRE06,*SegalPRL08,LindenPRL10,GemmerBOOK10}. On the  practical side, progress in computer nanotechnology is currently constrained by the need to understand and optimize power and cooling generation on space-\cite{Popieee06} and time- \cite{ErezNAT08,*GordonNJP09,*RaoPRA11,JahnkeEPL10} scales where quantum effects are unavoidable. This prompts
 the  strive to achieve a better grasp on the fundamental thermodynamic bounds of quantum devices, in particular, those that may act as heat machines. Quantum devices that convert information into work are outside the scope of our analysis.

 In traditional thermodynamics, the Carnot cycle \cite{CallenBOOK85} yields the ideal performance bound of heat engines and pumps (refrigerators). In this cycle the evolution consists of ``strokes'' in which the system (``working fluid'') alternates between coupling  to the ``hot'' and ``cold'' heat baths. Yet in microscopic or nanoscopic devices, certainly when they operate quantum mechanically, such cycles pose a serious problem: on-off switching of system-bath interactions may strongly affect energy and entropy exchange, which casts doubts on the validity of commonly discussed models that ignore such effects. \cite{CallenBOOK85, GeusicPR67,*GevaJMO02,*QuanPRE07,*VandenBroeckPRL05,*LinPRE03,*BenderJPA00,*EspositoPRL09,*QuanPRE07,AlickiOSID04,*AlickiJPA79,SegalPRE06,*SegalPRL08,LindenPRL10} 
 
 In this article we put forward a more rigorous approach to working cycles in quantum devices: we describe the steady-state dynamics of \textit{periodically-driven} open quantum systems that are \textit{permanently} coupled to heat baths by Floquet (harmonic) expansion of their coarse-grained Liouvillian evolution. The accuracy of this approach and its consistency with thermodynamics are ensured for weak system-bath coupling \cite{AlickiARXIV12}.
 Here we apply this theoretical machinery to the description and performance analysis of  a \textit{minimal model} of a quantum heat machine (QHM),  with the following features:
1) It is self-contained, i.e. described by a quantum mechanical Hamiltonian. 2) It is universal, i.e., it can act \textit{``on demand''}  as either a quantum heat engine (QHE) \cite{BlickleNATP11,*ChanNAT11,SegalPRE06,*SegalPRL08,LindenPRL10} that produces work, or as a quantum refrigerator (QR) \cite{gevajcp92,*FeldmannAJP96,*rezekNJP06,*kosloffJAP00,*FeldmannEPL10, LevyPRL2012,*cleurenPRL12,*mariPRL12}  that refrigerates a bath with finite heat-capacity, depending on a control parameter. 3) It is  broadly adaptable to  the available  bath/environment or temperature.
\par
Our  minimal model  consists of a \textit{single qubit} permanently  attached to both baths and controlled by a harmonic-oscillator ``piston''. A related (mostly numerical) study of QR has employed a qubit that alternately couples to one or another of two baths, whose spectra have different cutoffs \cite{SegalPRE06}.  A QR based on a harmonic oscillator that alternately couples to one of two qubits, each attached to a different Markovian bath, has also been suggested \cite{LindenPRL10}. 
Our goal  is the development, from first principles,  of a comprehensive analytical theory for universal,  dual-purpose (QHE or QR)  operation and its performance bounds in our minimal QHM model, wherein the  \textit{spectral  separation of the two baths  plays a key role.} Broad applicability is here ensured by \textit{bath engineering}: attaching a ``doorway mode'' to an \textit{arbitrary} bath acts as a \textit{ bandpass filter} that can impose the required spectral separation on the two baths. Such doorway  (filter)modes are realizable by interfacing the system (qubit) with the baths through a tunable cavity \cite{ViolaPRA98} or an impurity/defect in a periodic structure/chain \cite{KofmanJMO94}. 

The simplicity of the model allows closed-form analytical solutions, in which the piston-qubit \textit{coupling strength} is the ``knob'' that can transform a QR to QHE  (or vice versa) and controls their efficiency, after appropriately engineering the baths in question. Remarkably, Carnot efficiency is analytically shown to be achievable at the value of the control parameter that transforms the QR into a QHE. The optimal power and efficiency for finite-time cycles are analytically shown to surpass the established Curzon-Ahlborn bound \cite{CurzonAJP75}.

In Sec. II the model and the analysis framework are introduced. In Sec. III the periodically modulated steady state is evaluated by the Floquet expansion method. In Sec. IV the steady-state thermodynamic relations are derived. In Sec. V we investigate the operation modes and bounds of the QHM, based on the steady-state solution of Sec. IV. In Sec. VI we investigate the finite-time optimal performance bounds. Realizations are discussed in Sec. VII. The conclusions are presented in Sec. VIII.

{\section{Model and Treatment Principles}. 

The  Hamiltonian of the  QHM in question can be written as  
\begin{gather}
H_{QHM}=H_S(t)+H_B+H_{SB}; \\
H_{SB}=\sigma_x(B_H+B_C).
\end{gather}

Here the control two-level system (TLS) is weakly coupled simultaneously to two  baths via $H_{SB}$,
where $\sigma_x$ is the spinor x-component, $B_H$   and $B_C$ are respectively the operators  of a very large hot bath (H) and of a finite cold bath (C). The TLS frequency  is periodically modulated  about its  resonance  frequency $\omega_0$ by  the Hamiltonian 
 \begin{equation}
H_{S}(t)=\frac{1}{2}\sigma_z \nu(t)
\label{eq:tlsh}
\end{equation}

This model Hamiltonian is realizable by adiabatically eliminating  a highly detuned level of a three-level system and allowing for a periodic AC Stark  shift  by a time-dependent control field (Sec. VIIc). The fully quantized version of this model, wherein the classical time-dependent control field is replaced by a quantum  harmonic-oscillator field  dispersively coupled to   the TLS, merits separate discussion.

A scenario that illustrates the model (Fig. 1-inset) is as follows: A  charged quantum oscillator in a double-well potential which is ``sandwiched'' between the baths, a C-bath with finite heat-capacity and  a nearly-infinite H-bath which  serves as  heat dump. The oscillation is periodically modulated, e.g. by  off-resonant $\pi$-pulses, . These phase-flips control the heat current between the baths via the particle. 

 In the refrigerator mode, corresponding to   C-bath cooling, this model is reminiscent of the so-called sideband cooling: an optical Raman process  in solids and molecules \cite{Pringsheimzp129,*LandauJP46,*dousmanisapl63,*DjeuPRL81,*egorovjcp95,*LloydPRA97,*EpsteinNAT95,munganPRL97,*MunganJOSAB03}. Here  the red- and blue- shifted  TLS frequencies play  the role of Stokes and anti-Stokes lines of sideband cooling respectively: heat is pumped into an upshifted line in the H-bath spectrum, at the expense of a downshifted C-bath spectral line, the energy difference is supplied by the modulation.   In the engine mode, the opposite occurs: the  modulation converts part of the heat-flow energy from the H-bath to the C-bath into work extractable by the control field. This entails energy transfer from the H-bath to the field.

\begin{figure}
	\centering
		\includegraphics[width=0.55\textwidth]{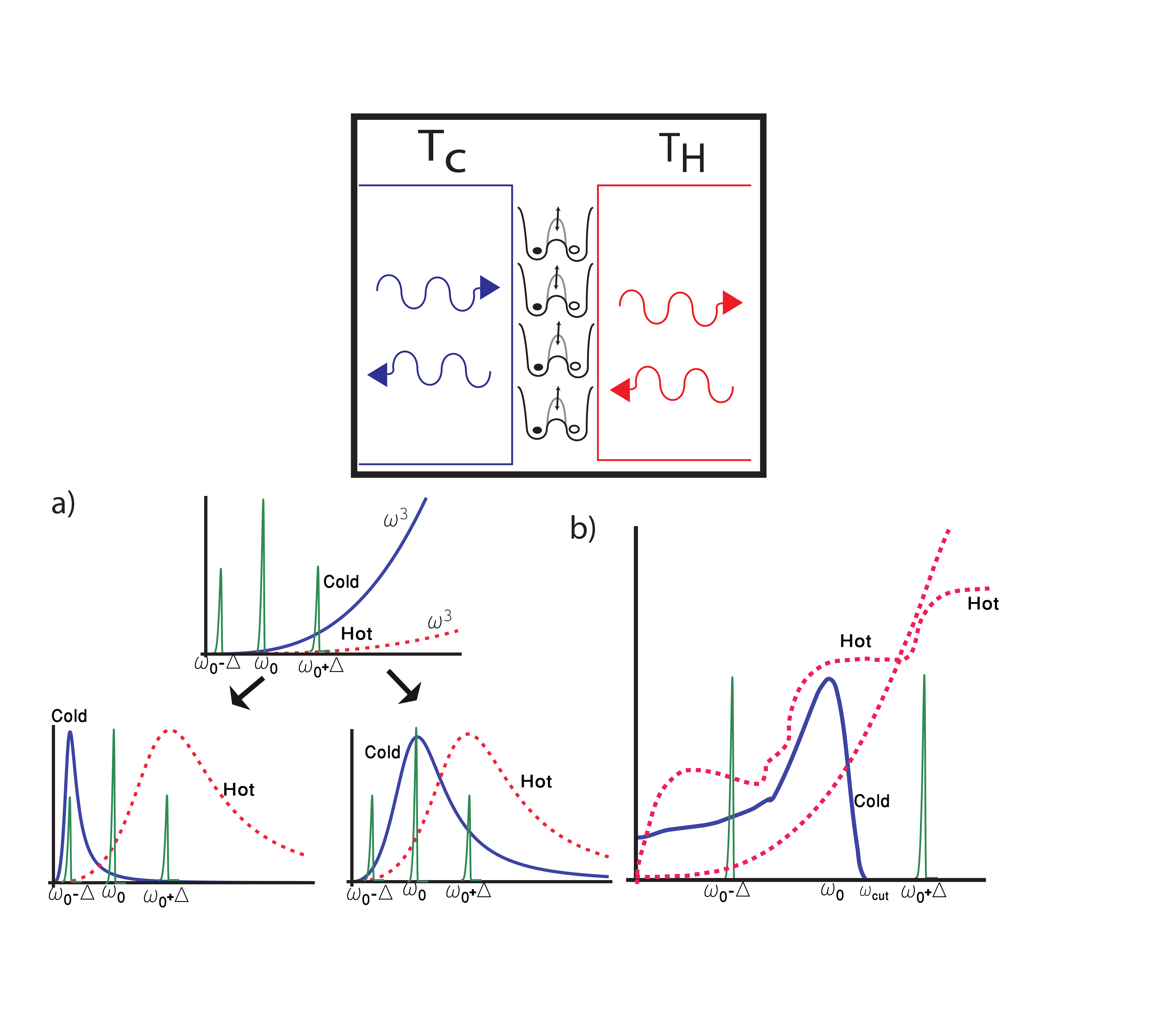}
	\caption{(color online) Inset: Illustration of the QHM by double-well qubits (with periodically-modulated tunneling barrier) embedded between cold and hot baths. a) Sinusoidal-modulation effects (Eqs. \eqref{ham_uni},\eqref{ham_uni1}). Harmonic (Floquet) peaks of the response superimposed on rising cold and hot bath spectra $G^j(\omega)=A^j\omega^3$(top); and on the spectra of the \textit{same baths} recalculated in the presence of different filter modes,  transforming these baths spectra into skewed Lorentzians (according to Eqs. \eqref{eq:efeg},\eqref{eq:DElta}). The filtered spectra obey condition A1 (left) or A2 (right). The parameters (in arbitrary units) used to calculate the spectra are: $A^C=1,A^H=1/10$, for the graph on the left:$\gamma_f^H=22$, $\gamma_f^C=1$, $\omega_f^H=13$ and $\omega_f^C=1$ and on the right $\gamma_f^H=1$, $\gamma_f^C=2$, $\omega_f^H=13$ and $\omega_f^C=20$.  b) Same, under condition B (phase-flip modulation), for rising hot-bath spectra and cold-bath spectrum with cutoff. }
	\label{fig:modfig}
\end{figure}

\begin{figure}
\hspace{-1cm}
	\centering
		\includegraphics[width=0.5\textwidth]{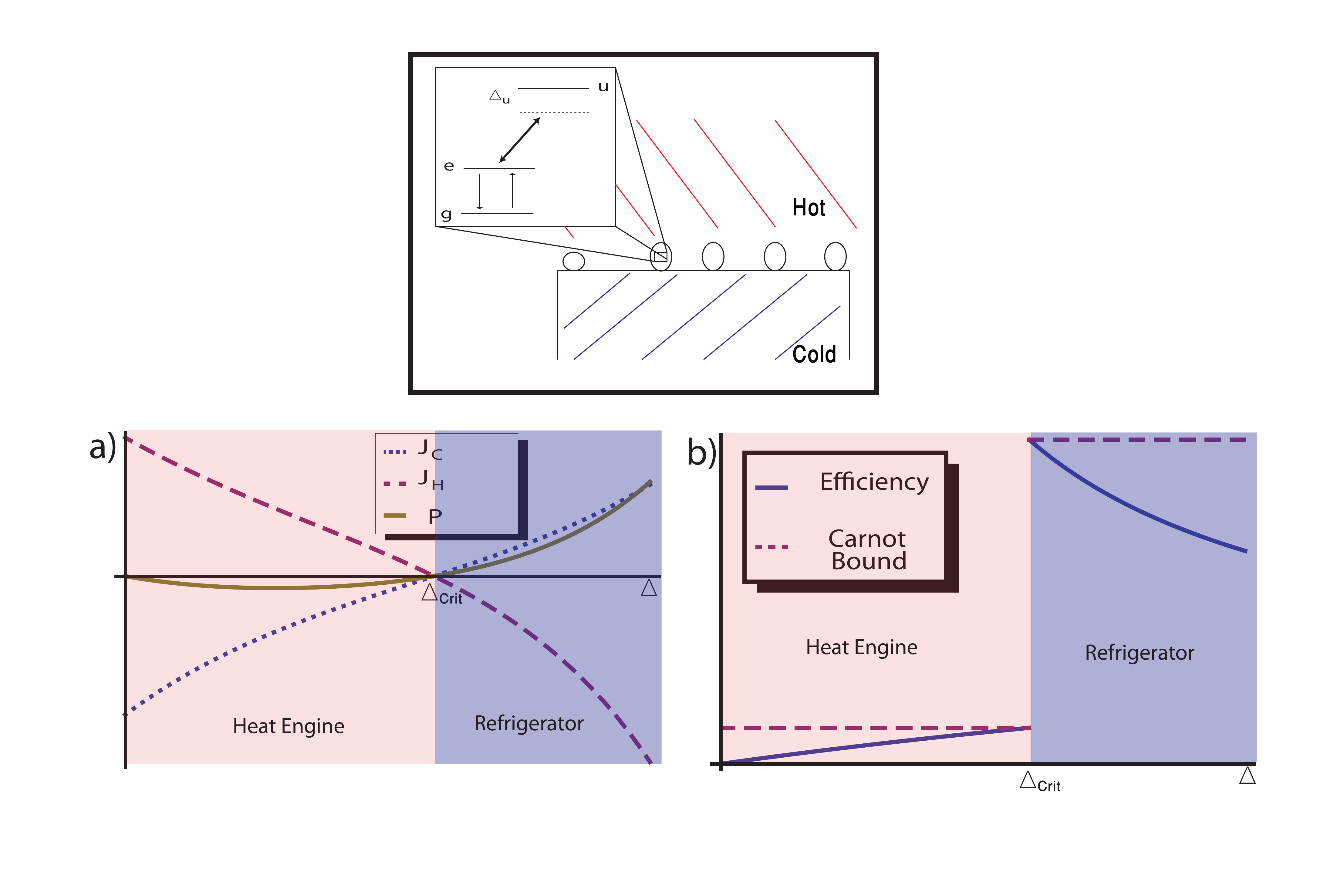}
	\caption{(color online) a) Currents and power as function of the machine modulation for an ideal heat machine. For $\Delta <\Delta_{cr}$ it operates as a heat engine and for $\Delta >\Delta_{cr}$ as a refrigerator. b) Efficiency for the same machine. The Carnot bound is reached at $\Delta=\Delta_{cr}$.  Inset: Periodic modulation of 3-level impurities embedded between hot and cold baths. The modulating field is detuned from the $|e\rangle-|u\rangle$ transition by $\Delta_u$ and  exerts periodic AC Stark shift on $|e\rangle$.}
	\label{fig:impurity2}
\end{figure}

 Work-extraction or refrigeration conditions are determined by  the  direction of power and heat flow (the  heat current). The heat current, in turn, is given by the polarization rate of the TLS, obtained from the  steady-state solution of a  master equation (ME) for the TLS density operator.  This ME, which allows for non-Markovian (bath-memory) effects,  is accurate to second order in the system-bath coupling,  at any temperature, as verified by us  both theoretically \cite{KofmanPRL04,*GordonJPB07,*GordonPRL08} and experimentally \cite{almogjpb11,AlvarezPRL10}.  Inaccuracies of the ME\cite{GordonNJP10}  are  negligible for weak coupling. (See Suppl. Mat. \cite{KolarPRL12}) 
 
 The refrigeration  of a finite-capacity C-bath represents a succession  of tiny temperature changes over many modulation cycles. Hence, the Born approximation underlying the ME is consistent with such cooling. The finite-capacity C-bath is assumed to have  a continuous spectrum, since bath-mode discreteness and the associated  recurrences may render the Born approximation invalid and preclude bath thermalization \cite{smithephb08}.
 
The TLS density matrix $\rho_S$  is assumed diagonal in the energy-state basis: starting at equilibrium, where off-diagonal elements of $\rho_S$  are   absent, they  remain so, when the TLS is subject to $\sigma_Z-$ modulation in Eq. (2). The  interlevel transition rates and  their non-Markovian time-dependence   embody the quantumness of the ME.
Sufficiently fast periodic modulation of the TLS frequency  at intervals $\tau$ can change  the detailed balance of the transition rates (quanta absorption and emission) and thereby  allows either to heat up or cool down the TLS  depending on $\tau$ \cite{ErezNAT08,AlvarezPRL10}.
Here  our goal  is the analysis of   work extraction or  heat flow between the \textit{baths} enabled by   periodic modulation  of the TLS. Two alternative methods  yield the same equations for these processes: a) Floquet (harmonic) expansion of the non-Markovian ME \cite{ErezNAT08,KofmanPRL04} under  temporal averaging (coarse-graining over a modulation period); b) Floquet expansion of the  Markovian evolution superoperator \cite{AlickiARXIV12,AlickiPRA06}.

\section{ Floquet expansion at  steady state}
\subsection{Non-Markovian master-equation approach}
Since  $\rho_S$ is diagonal (see above) in the energy basis of the TLS ($|e\rangle$, $|g\rangle$), the diagonal $H_S(t)$ (Eq. \eqref{eq:tlsh}) yields the following rate equations from the non-Markovian ME \cite{KofmanPRL04}
\begin{equation}
\dot\rho_{ee}(t) = -\dot\rho_{gg}(t) = R_g(t)\rho_{gg}-R_e(t)\rho_{ee},
\label{eq-Bloch}
\end{equation}

The non-Markovian, time-dependent $|e\rangle \rightarrow |g\rangle$ and $|g\rangle \rightarrow |e\rangle$ transition rates are given , respectively, by the real part of the integrals ~\cite{KofmanPRL04,KofmanPRL01,*KofmanNAT00}
\begin{multline}
R_e(t)=
2\,{\rm Re}\!\int_0^t{\rm d}t^\prime\exp[i\omega_0(t-t^\prime)]\varepsilon(t)\varepsilon^\star(t^\prime)\Phi(t-t^\prime),\label{rates-def}\\
R_g(t)=2\,{\rm Re}\!\int_0^t{\rm d}t^\prime\exp[-i\omega_0(t-t^\prime)]\varepsilon^\star(t)\varepsilon(t^\prime)\Phi(t-t^\prime),
\end{multline}
where  $\varepsilon(t)$ is  the periodically modulated phase factor (a unimodular periodic complex function),  $\omega_0$ is the TLS resonance frequency and the bath-response (autocorrelation) function $\Phi(t)\equiv \int {\rm d} \omega G(\omega)\exp(-i\omega t)$ is the  Fourier transform of to  the bath coupling spectrum $G_T(\omega)$.

The TLS  evolution caused by  the baths under weak coupling  conditions is much slower than their memory (correlation) time $t_c$. Hence, in steady state,  we can use  \textit{time-averaged (coarse-grained)  level populations and transition rates} (See Suppl. Mat. \cite{KolarPRL12}.
In Eq.\eqref{eq:tlsh}  the coarse-grained   dynamics yields the following \textit{additive} contributions of the two baths (labeled by $j=C,H$) to the harmonic expansion  (labeled by m) of the  time-averaged TLS polarization $\overline{S}\equiv \frac{\overline{\rho_{ee}-\rho_{gg}}}{2} $
\begin{gather}
\dot{\overline{S}}=\sum_m(\dot{\overline{S^C_m}}+\dot{\overline{S^H_m}}) \nonumber\\
-\sum_{m,j} \left(
-(\overline{R_g^{j(m)}}+\overline{R_e^{j(m)}})\overline{S}+\frac{\overline{R_g^{j(m)}}-\overline{R_e^{j(m)}}}{2} \right),
\label{eq:sdot}
\end{gather}

where the time-averaged transition rates  are found from the Floquet expansion of the 
 modulation $\nu(t)$ to be\cite{KofmanPRL04,CallenBOOK85,KofmanPRL01}

\begin{equation}
\overline{R}^j_{e(g)}\equiv 
2\pi\sum_m P_m G^j[\pm(\omega_0+m\Delta)];
\label{decay-rates-averaged-def}
\end{equation}

Here 
\begin{equation}
P_m = |\varepsilon_m|^2\  , \ \varepsilon_m = \frac{1}{\tau}\int_0^{\tau} e^{i\int_0^t(\nu(t')-\omega_0)dt' }e^{im\Delta t} dt.
\label{eq:}
\end{equation}

\noindent  are the  probabilities of shifting  the j-th-bath coupling spectrum $G^j(\omega)$  by $m\Delta$, $\Delta=\frac{2\pi}{\tau}$, from the  average frequency $\omega_0$, and
\begin{equation}
G^j(\omega)= \int_{-\infty}^{+\infty} e^{i\omega t}\langle B_j(t)B_j(0)\rangle dt = e^{\omega/ T} G^j(-\omega) 
\label{eq:Gk}
\end{equation}

For a bosonic bath ($\hbar=k_B=1$):

\begin{eqnarray}
G^j(\omega)=&G^j_0(\omega)(n_j(\omega)+1);  \hspace{0.5cm}  G^j_0(\omega)=|g_j(\omega)|^2\rho_j(\omega);
   \label{eq:G}
 \end{eqnarray}
 $g_j(\omega)$ being the system-bath coupling, $\rho_j(\omega)$ the bath -mode density and $n_j(\omega)=\frac{1}{e^{\frac{\omega}{T_j}}-1})$ the $\omega$-mode thermal occupancy.

\subsection{Markovian master equation approach}

An alternative method is based on the combination of the weak-coupling limit and Floquet expansion \cite{AlickiPRA06} of a periodically-flipped qubit coupled to two baths is based on the Lindblad-Gorini-Kossakowski-Sudarshan (LGKS) operator, as expounded in the tutorial in Ref. \cite{AlickiARXIV12}. It has the advantage of ensuring positivity and additivity of the evolution due to the two baths.  The Markovian master equation in the interaction picture reads: $\frac{d\rho}{dt}=\mathcal{L} \rho$ where  $\mathcal{L}_{j,m}$ we expand $\mathcal{L=\sum L}_{m}^{j},$ where again $j=H,C$, and m is the Floquet harmonic. The expansion yields  

\begin{gather}
\mathcal{L}_{m}^{j}\rho=\frac{P_m}{2}\Bigl(G^{j}(\omega_{0}+m\Delta)\bigl([\sigma^{-}\rho,\sigma^{+}]+[\sigma^{-},\rho\sigma^{+}]\bigr)+\notag \\
G^{j}(-\omega_{0}-m\Delta)\bigl([\sigma^{+}\rho,\sigma^{-}]+[\sigma^{+},\rho\sigma^{-}]\bigr)\Bigr)
\label{gen_qubit2}
\end{gather}

This (LGKS) approach  yields the same equations of motion for the qubit polarization  as the \textit{time-averaged} non-Markovian approach above (Eq. \eqref{rates-def}-\eqref{eq:sdot}).

\section{Steady-State thermodynamic relations}

The qubit steady-state (found from Eq.\eqref{gen_qubit2})   has the diagonal  form characterized by the population ratio $w$

 $\tilde{\rho}=\left(\begin{array}{cc}
\tilde{\rho_{ee}} & 0\\
0 & \tilde{\rho_{gg}}
\end{array}\right)$ ,

\begin{equation}
w=\frac{\tilde{\rho_{ee}}}{\tilde{\rho_{gg}}}=\frac{\sum_{q,j}P_m G^{j}(\omega_{0}+m\Delta)e^{-\frac{\omega_{0}+m\Delta}{T_{j}}}}{\sum_{m,j}P_m G^{j}(\omega_{0}+m\Delta)}
\label{eq:ss}
\end{equation}

where, as in Eq.\eqref{eq:sdot}, $\Delta=\frac{2\pi}{\tau}$. The cold (hot) current is then given by

\begin{equation}
J_{C(H)}=\sum_{m}(\omega_{0}+m\Delta)P_mG^{C(H)}(\omega_{0}+m\Delta)\frac{e^{-\frac{(\omega_{0}+m\Delta)}{T_{C(H)}}}-w}{w+1}
\label{eq:curr}
\end{equation}

 The magnitudes and signs of these steady-state currents are the same as for the time-averaged (non-Markovian) ME solutions detailed above (Eq. \eqref{rates-def}-\eqref{eq:sdot}).

The first law of thermodynamics allows to define the stationary power as
\begin{equation}
\mathcal{P} = -( J_{C} +J_{H} ).
\label{power}
\end{equation}
i.e. the  power investment $\mathcal{P}$ by the piston  is negative when the machine acts as an engine. 
One obtains from \eqref{power}
\begin{equation}
\mathcal{P}= \sum_{m,j}\frac{(\omega_0+m\Delta)P_m}{w+1}\left[G^{j}(\omega_0+m\Delta)(w-e^{-\frac{(\omega_0+m\Delta)}{T_{j}}})\right].
\label{power_gen}
\end{equation}

 One can show using the Spohn theorem for Markovian evolution \cite{SpohnJMP78} that the  heat currents satisfy the second law of thermodynamics 
\begin{equation}
 \mathcal{S}(t)= -\mathrm{Tr}\bigl(\tilde{\rho}(t)\ln \tilde{\rho}(t)\bigr) \   ,\
 \frac{d}{dt}\mathcal{S}(t)- \frac{J_{C}(t)}{T_C} + \frac{J_{H}(t)}{T_H} \geq0
\label{II_law}
\end{equation}
where the left hand side of the inequality (\ref{II_law}) is the entropy production rate.
\par

These standard thermodynamical relations imply the validity of the Carnot bound on the engine efficiency $\eta$, and  on the coefficient of performance (COP) for the refrigerator
\begin{gather}
\eta = \frac{-\mathcal{P}}{J_{H}}\leq 1-\frac{T_C}{T_H},\  \nonumber \\
COP =\frac{J_{C}}{\mathcal{P}}\leq \frac{T_C}{T_H - T_C}.
\label{efficiency}
\end{gather}

\section{Universal Machine Operation Modes}
The choice of parameters that may affect the QHM operation are mainly  the shape of the modulation $\nu(t)$ and the form of bath-response (coupling spectra) spectral densities $G^C(\omega)$ and $G^H(\omega)$. These choices may enable the machine to  act  as both an engine and a refrigerator, as shown below. In the following we discuss two such choices of the modulation and the requirements each type entails on the bath spectra. 

A. \textbf{Sinusoidal modulation}

 We consider  the sinusoidal time-dependence of the external  (modulating) field, i.e.
\begin{equation}
\omega(t) = \omega_0 + \lambda \Delta\sin(\Delta t) 
\label{ham_uni}
\end{equation}
under the condition
\begin{equation}
0\leq \lambda << 1. 
\label{ham_uni1}
\end{equation}
The condition (\ref{ham_uni1}) implies that only the harmonics $m=0,\pm1$ with 
\begin{equation}
P_{m=0} \simeq 1- \frac{\lambda^2}{2}\ ,\ P_{m=\pm1} \simeq \frac{\lambda^2}{4}
\label{ham_uni2}
\end{equation}
should be taken into account. Even under this simplifying condition the formulae for heat currents and power are  complicated (see Appendix A). More detailed analysis of the various terms in those formulae shows that
in order to reach the Carnot bound, we have to reduce the number of relevant harmonics to two. This can be done by  system - bath coupling engineering (Sec. VII), so as  to impose \emph{the spectral separation of the two baths}, in the cases discussed below.

\textit{A1} 
We assume that the upper cutoff of C nearly coincides with lower cutoff of H and $\omega_0$ is near the two cutoffs(Fig. 1a):
\begin{equation}
G^C(\omega)\simeq 0 \ \mathrm{for}\  \omega\geq \omega_0\ \mathrm,\  G^H(\omega)\simeq 0\  \mathrm{for}\ \omega \leq \omega_0 , 
\label{spectral_sepA}
\end{equation}

\par

\par 
\noindent We discuss this case in more detail, as  the condition (\ref{spectral_sepA}) is the  easiest to implement in practice (see Section \ref{sec:rea}).
This condition leads to the following simplified formulae for heat currents and power obtained from the general expressions (Appendix A)
\begin{eqnarray}
J_{H}&=&(\omega_0+\Delta)\mathscr{N}(e^{-(\frac{\omega_0+\Delta}{T_{H}})}-e^{-(\frac{\omega_0-\Delta}{T_{C}})}),\nonumber\\
J_{C}&=& -(\omega_0-\Delta)\mathscr{N}(e^{-(\frac{\omega_0+\Delta}{T_{H}})}-e^{-(\frac{\omega_0-\Delta}{T_{C}})}),\nonumber\\
\mathcal{P}&=&-2\Delta \mathscr{N}(e^{-(\frac{\omega_0+\Delta}{T_{H}})}-e^{-(\frac{\omega_0-\Delta}{T_{C}})}),
\label{current_uni}
\end{eqnarray}
where  the \textit{positive} normalizing constant is 
\begin{gather}
\mathscr{N} = \nonumber \\
 \frac{\lambda^2}{4}\frac{G^{C}(\omega_0-\Delta) G^{H}(\omega_0+\Delta)}{G^{C}(\omega_0-\Delta)\bigl[1 + e^{-(\frac{\omega_0-\Delta}{T_C})}\bigr] +G^{H}(\omega_0+\Delta)\bigl[1 + e^{-(\frac{\omega_0+\Delta}{T_H})}\bigr]}.
\label{N}
\end{gather}
It follows from (\ref{current_uni}) that there exists a critical value of the modulation frequency
\begin{equation}
\Delta_{cr} =\omega_{0}\frac{T_{H}-T_{C}}{T_{H}+T_{C}},
\label{critical}
\end{equation}
such that for $\Delta < \Delta_{cr}$ the machine acts as an engine with the efficiency
\begin{equation}
\eta=\frac{2\Delta}{\omega_{0}+\Delta},
\label{eq:efficiency}
\end{equation}
and for $\Delta > \Delta_{cr}$ as a refrigerator with
\begin{equation}
COP=\frac{\omega_{0}-\Delta}{2\Delta}.
\label{COP}
\end{equation}
At $\Delta = \Delta_{cr}$ the engine/refrigerator reaches its maximal Carnot efficiency/COP. This corresponds to  the  vanishing value of power/cold current (see Fig. \ref{fig:impurity2}a). The operation mode change prevents the  machine from breaking the second law. Similar behavior  was obtained  \cite{BirjukovEPJB08} by numerical calculation for a different model of a quantum machine.

\textit{A2} We assume the following conditions of the bath spectra (Fig. \ref{fig:modfig} a)%
\begin{equation}
G^C(\omega)\simeq 0 \ \mathrm{for}\  \omega \approx \omega_0 \pm \Delta \ \mathrm,\  G^H(\omega)\simeq 0\  \mathrm{for}\ \omega \leq \omega_0. 
\label{spectral_sepA1}
\end{equation}
This condition yields (App. A)

\par 
\begin{eqnarray}
J_{H}&=&(\omega_0+\Delta)\mathscr{N'}(e^{-(\frac{\omega_0+\Delta}{T_{H}})}-e^{-(\frac{\omega_0}{T_{C}})})\nonumber\\
J_{C}&=& \omega_0\mathscr{N'}(e^{-(\frac{\omega_0}{T_{c}})}-e^{-(\frac{\omega_0+\Delta}{T_{H}})})\nonumber\\
\mathcal{P}&=&-\Delta \mathscr{N'}(e^{-(\frac{\omega_0+\Delta}{T_{H}})}-e^{-(\frac{\omega_0}{T_{C}})}),
\label{current_uni1}
\end{eqnarray}
where   the \textit{positive} normalizing constant for this case is 

\begin{gather}
\mathscr{N'} =  \nonumber\\
\frac{\lambda^2}{4}\frac{G^{C}(\omega_0) G^{H}(\omega_0+\Delta)}{G^{C}(\omega_0)\bigl[1 + e^{-(\frac{\omega_0}{T_C})}\bigr] +\frac{\lambda^2}{4}G^{H}(\omega_0+\Delta)\bigl[1 + e^{-(\frac{\omega_0+\Delta}{T_H})}\bigr]}.
\label{N'}
\end{gather}

From Eq. \eqref{current_uni1} it follows that the critical modulation frequency is

\begin{equation}
\Delta_{cr} =\omega_{0}\frac{T_{H}-T_{C}}{T_{C}}.
\label{critical1}
\end{equation}
Namely  for $\Delta < \Delta_{cr}$ the machine acts as an engine with the efficiency
\begin{equation}
\eta=\frac{\Delta}{\omega_{0}+\Delta},
\end{equation}
and for $\Delta > \Delta_{cr}$ as a refrigerator with
\begin{equation}
COP=\frac{\omega_{0}-\Delta}{\Delta}.
\label{COP1}
\end{equation}
At $\Delta = \Delta_{cr}$ the engine/refrigerator reaches its maximal Carnot efficiency/COP.

\textbf{B $\pi$-flips modulation}

Periodic $\pi$-phase shifts (phase flips) with alternating sign  give   rise, to only two leading harmonics, corresponding to two symmetrically-opposite frequency shifts  in the Floquet expansion of the probability distribution (since $P_0=0$ for symmetry reasons)\cite{KofmanPRL04,KofmanPRL01}. 

  \begin{gather}
P_{\pm 1}\approx (2/\pi)^2 \notag;\\
G^j(\omega_0) \rightarrow G^j(\omega_0\pm\Delta)
\label{eq:p}
\end{gather}

If $\Delta$, is comparable to  $1/t_c$,  the inverse memory time of the cold bath, we may require that  at $\omega \simeq \omega_0+\Delta$ the TLS be coupled only to the H bath, while at $\omega\simeq\omega_0-\Delta$ it is  coupled to both the C  and the H bath. This is tantamount to the requirement that
 
\begin{eqnarray}
G^H(\omega_0+\Delta)\gg G^C(\omega_0+\Delta),G^H(\omega_0-\Delta), G^C(\omega_0-\Delta)
\label{eq:gt}
\end{eqnarray}

This requirement can be satisfied if $C$ has an upper cutoff 

\begin{equation}
\omega_{\rm cut}<\omega_0+\Delta. 
\label{eq:wcut}
\end{equation}

For $H$, by contrast,  $G^H$ is only required to rapidly rise with $\omega$, which is true  for blackbody radiation in open space, $G^H(\omega)\propto \omega^3$, or for phonons in bulk media.

If   Eq. \eqref{eq:gt} is satisfied  we find  the same steady-state expression for the currents and power   and the same  physical behavior as for case A1: $\Delta_{cr}$ is given by  Eq. \eqref{critical}, $\eta$   by Eq. \eqref{eq:efficiency} and COP by Eq. \eqref{COP}.

\section{Finite-time optimal performance}
The  vicinity  of the critical frequency is not a useful working regime, since power and currents are small there.
Much more important is the region of parameters where the  power or cold current are maximal. We wish to find these parameters and the corresponding efficiency of our QHM, and compare it to the  so-called Curzon-Ahlborn efficiency  at maximum power of a macroscopic Carnot-type engine\cite{CurzonAJP75}
\begin{equation}
\eta_{CA}= 1 -  \Bigl(\frac{T_{C}}{T_{H}}\Bigr)^{1/2}
\label{CAo}
\end{equation}
To this end we consider the case A2  (Eq. \eqref{spectral_sepA1}) for our universal machine and compute the maximal power $\mathcal{P}$ which is produced for the optimal modulation frequency $\Delta_{max}$ under the following simplifying assumptions.\\
a) The spectral density $G^{H}(\omega)$ is flat, i.e. $\frac{d}{d\omega}G^{H}\simeq 0$ near  $\omega=\omega_0 +\Delta_{max} $ .\\
b) high-temperature regime $e^{\frac{\omega}{T_a}}\approx 1+ \frac{\omega}{T_a}$. 

Under these assumptions, the modulation $\Delta_{max}$ that yields the  maximal power and the corresponding efficiency $\eta_{max}$ are found to be
\begin{equation}
\Delta_{max} = \frac{1}{2}\Delta_{cr}\ ,\ \eta_{max}= \frac {( 1 - \frac{T_C}{T_H})}{1+ \frac{T_C}{T_H}} \geq \eta_{CA}
\label{CA}
\end{equation}
\begin{center}
\begin{table*}[ht]
{\small
\hfill{}
\begin{tabular}{|c|c|c|c|c|c|}
\hline 
Case & $\Delta_{cr}=2\Delta_{max}$ & Efficiency ($\Delta \leq \Delta_{cr}$)&  Efficiency at maximum power & Relation to Curzon-Ahlborn\tabularnewline
\hline 
\hline 
A1 & $\omega_{0}\frac{T_{H}-T_{C}}{T_{H}+T_{C}}$  & $\frac{2\Delta}{\omega_{0} + \Delta}$ & $\eta_{max}=\frac{2(T_H - T_{C})}{3 T_H +T_{C}}$ & $\eta_{max}\leq \eta_{CA}$ \tabularnewline
\hline
A2 & $\omega_{0}\frac{T_{H}-T_{C}}{T_{C}}$ & $\frac{\Delta}{\omega_{0}+\Delta}$  & $\eta_{max}=\frac{T_{H}-T_{C}}{T_{H}+T_{C}}$ & $\eta_{max}> \eta_{CA}$ \tabularnewline
\hline 
\end{tabular}}
\hfill{}
\end{table*}
\end{center}

In the Table 1 we present the results of  cases A1 (Eq. \eqref{spectral_sepA}) and A2 (Eq. \eqref{spectral_sepA1}). While the relation  $\Delta_{max} = \frac{1}{2}\Delta_{cr}$ is true for all cases, at maximal-power  the efficiencies are different.
In the case A2 the Curzon-Ahlborn bound is always exceeded. In fact,  $\eta_{CA}$  
is then  the minimum of $(\eta)_{max}$.
Numerical calculation shows that the Curzon-Ahlborn bound is exceeded
even if the  temperatures are not high ($\Delta\approx\omega_0/2,T_{H}\simeq2,T_{C}\simeq1)$.
\section{Realization considerations} \label{sec:rea}
In the following we lay out general guidelines for experimental realizations i.e. the key considerations for satisfying the operating conditions discussed above:

a)\textit{Spectral separation of coupling to the C and H baths via Debye cutoffs} 

Case B (periodic $\pi$-flips) is the most flexible so far as spectral separation is concerned: it allows for an arbitrary rising coupling spectrum of the H bath, including a bulk-solid phonon bath or the blackbody radiation  spectrum $G^H(\omega)\propto \omega^3$. The C bath should have $G^C(\omega=0)\neq 0$, as for  1/f noise spectra \cite{ClausenPRL10,*clausenPRA12}. Preferably, the C bath is to have an upper cutoff, as is the case for phonon baths in crystals (Debye cutoff) and Ohmic or super-Ohmic noise baths\cite{ClausenPRL10}. By contrast, a single mode of a cavity with finite linewidth (finesse) is a Lorentzian bath\cite{KofmanPRA96}, whose lack of cutoff would lower the machine efficiency or COP, unless the upshifted (anti-Stokes) frequency $\omega_0 + \Delta$   is at the far tail of the Lorentzian (Fig. 1).

The spectral separation requirements are stricter in cases A1 and A2 (under sinusoidal modulation). To impose these requirements, 
we can choose the materials C and  H in such a way that the Debye frequencies match the desired temperatures of the baths 
\begin{equation}
\omega_D^C \simeq T_C\ ,\ \omega_D^H \simeq T_H.
\label{match}
\end{equation}

The coupling  spectra of the baths are assumed to have the Debye shape 
\begin{equation}
G^j(\omega) =  f_j\Bigl(\frac{\omega}{\omega_D^j}\Bigr)^3\frac{1}{1-e^{-\omega/T_j}} \theta(\omega_D^j-|\omega|)
\label{match1}
\end{equation}

where $f_j$ are bath-specific constants and $\theta$ is the Heaviside step function.

b) \textit{Filters} 

In general, the satisfaction of the spectral separation conditions can be facilitated by imposing a ``filter'' onto the qubit-bath coupling spectrum. To this end, we consider the model whereby the qubit is (weakly) coupled to two harmonic-oscillator ``filter''modes with resonance frequencies $\omega_f$ each such mode in turn, is coupled to the respective (C or H) bath via coupling spectrum $G^j(\omega)$, i.e. each filter mode mediates between the qubit and the respective bath. The qubit is then effectively coupled to the filter-mode bath response via\cite{KofmanJMO94}

\begin{equation}
G_f^j(\omega)=
\frac{\gamma_f}{\pi}
\frac{(\pi G^j(\omega))^2}
{(\omega-(\omega_f^j+\Delta_L^j(\omega)))^2+
(\pi G^j(\omega))^2},
\label{eq:efeg}
\end{equation}
 
 where $\gamma_f^j$ is the coupling rate of the qubit to the filter mode, and 
 
 \begin{equation}
 \Delta_L^j(\omega)=P(\int_0^{\infty} d\omega' \frac{G^j(\omega')}{\omega-\omega'})
\label{eq:DElta}
\end{equation}
 P being the principal value, is the respective bath-induced Lamb shift \cite{KofmanJMO94,cohentBOOK97}. The filter-mode response spectrum \eqref{eq:efeg} is a ``skewed Lorentzian'' for a completely \textit{general} spectrally structured $G^j(\omega)$. In particular, a cutoff or a bandedge of  $G^j(\omega)$ curtails the Lorentzian and makes it strongly skewed \cite{KofmanJMO94}, whereas if  $G^j(\omega)$ is spectrally flat, it is a simple Lorentzian centered at $\omega_f^j$. Such ``filtering'' can suppress  undesirable tails of  $G^j(\omega)$ and thus enforce conditions \eqref{spectral_sepA} or \eqref{eq:gt} in a broader range of media and parameters (Fig. 1a).
 
 In addition, the desired bath spectra are achievable by engineering. We note recent advances in microcavities, photonic crystals and waveguides that can help reshape photon bath spectra, as well as their phonon-bath counterparts, e.g., periodic structures with acoustic bandgaps \cite{ChanNAT11}.
 
c) \textit{Qubit realization and modulation}

A qubit may be realized by atomic or molecular multilevel impurities embedded at the interface between different material layers (Fig.2-inset)
A qubit may also be realized by a symmetric double-well (DW) potential with two bound states $|e\rangle$, $|g\rangle$: for example, a single-electron quantum dot  or a superconducting Josephson qubit\cite{RablPRL06,BenskyQIP11} as well as its ultracold-atom analog \cite{BargilPRA09,*BargilPRL11,*BaronePRL04}. The  symmetric and antisymmetric superpositions of $|e\rangle$ and $|g\rangle$, the eigenstates of $\sigma_x$ (Eq. (2)), are localized on the left- and right-hand well, respectively. In either bath the Debye cutoff should conform to the temperature as  specified above. One example is that of   a DW quantum dot ``sandwiched'' between dielectric layers with different Debye cutoffs $\omega_D^j$ (Fig.1-inset) Another example is an ultracold-atom DW qubit embedded between two optical lattices, where phonons  have different $\omega_D^j$ (Fig.2-inset). 

 The required \textit{modulation} that conforms to Eqs. (19) or (23) may be realized by changing the energy difference of $|e\rangle$ and $|g\rangle$ by time-dependent AC Stark shift\cite{KofmanPRL04}. If $|u\rangle$ is an upper state with energy $\omega_u$, then an off-resonant control field detuned by $\Delta_u$ from the $|e\rangle$-$|u\rangle$ resonance will realize the piston-system coupling in Eq. (3) with $\nu (t)=\Omega^2(t)/ \Delta_u$, $\Omega(t)$ being the Rabi frequency of the control field. (Fig. 2-inset).Thus, in DW qubits realized by ultracold atoms, the optically-induced potential barrier between the wells may be periodically modulated or flipped in sign \cite{BargilPRA09} (Fig. 1-inset)

In order to illustrate the performance of such a machine,   we  numerically compute the dependence of the basic thermodynamical parameters on the rate  $\Delta$.
As shown in  Fig. 2, the operation mode, as well as its efficiency, depends on $\Delta$.  In  the plotted  example, the overlap of spectral densities of the two baths  vanishes, i.e. they fully  satisfy the spectral separation condition A, hence the ideal Carnot efficiency is reached. If the overlap does not vanish, the Carnot bound is not attained.

\section{Conclusions}
A  single TLS (qubit) with energy modulation has been shown to constitute a \textit{minimal model}, for a universal quantum heat machine 
 (QHM) that is permanently attached to two spectrally separated baths. Such simultaneous coupling to both baths allows our rigorous analysis of energy and entropy exchange in a cycle. The present approach   stands in contrast to the traditional  cycles division ded into ``strokes'', each involving one bath at a time  where system-bath on-off switching effects are no accounted for. 
\cite{GeusicPR67,*GevaJMO02,*QuanPRE07,*VandenBroeckPRL05,*LinPRE03,*BenderJPA00,*EspositoPRL09,*QuanPRE07,AlickiOSID04,*AlickiJPA79,SegalPRE06,*SegalPRL08,LindenPRL10,GemmerBOOK10,Popieee06}.   As shown,  the analysed machine  can be switched at will from an engine mode to refrigerator mode and vice versa merely by varying the modulation rate. This could be  useful in situations where both kinds of thermal machines are needed, and the  operation is simplified by having a single machine.   The rate also allows  to control the efficiency of our machines and in this way keep them optimized under bath temperature change, which underscores the versatility of our machine. Recently we  have shown that this machine may \textit{violate the unattainability of the absolute zero} (the third law) in the refrigeration mode for certain bath models \cite{KolarPRL12}.

The switching from engine to refrigerator mode in this machine occurs at a critical rate,i.e. at the critical point where Carnot limit appears  to be close to be broken, yet instead the mode  switching prohibits breaking this limit. The maximum efficiency is reached at the critical point, where the machine yields zero power, consistently with the second law.
Practical engines are however designed to yield  maximum power. It is usually presumed that the upper bound is the Curzon-Ahlborn efficiency \cite{CurzonAJP75}. However, as our model shows,  this bound can be exceeded, it is in fact the  \textit{lower} bound for our finite-time engine.

Both impurities and double-well qubits embedded in appropriate environments may act, depending on their energy modulation, in either the engine (QHE) or the refrigerator (QR) mode of the proposed QHM, with potentially significant technological advantages. In particular, an impurity or quantum-dot fast-modulated qubit ``sandwiched'' between two nanosize solid layers (Fig.1-inset) may act as  a nanoscopic refrigerator (heat-pump) of a transistor (chip), that is much more miniaturized and less power-consuming than currently available microelectronic refrigerators \cite{Popieee06}. Under slower modulation rate, the same setup may act as electron-current generator \textit{without external voltage bias}, and a substitute for phase-coherent electron-current control\cite{KurizkiPRB89} (whereas in the present scheme the modulating field need not be coherent). 

To conclude, the present scheme demonstrates the ability of systematic quantum analysis of driven open systems to yield accurate, physically lucid, expressions for steady-state heat machine performance. The analysis shows that quantum mechanics and thermodynamics can be fully compatible on the level of an elementary (\textit{single-qubit}) system provided one \textit{correctly accounts} (by the Floquet expansion) for its entropy and energy exchange with the baths. This essential point has not been properly accounted for by previous treatments of quantum heat machines.

\textit{Acknowledgements} The support of ISF, DIP, the Humboldt-Meitner Award(G.K.), the Weston Visiting Professorship (R.A.), CONACYT (D.G.) are acknowledged.

\setcounter{equation}{0}
\renewcommand{\theequation}{A\arabic{equation}}

\subsection*{Appendix A. Heat currents and power under sinusoidal modulation}
In general, under the conditions\eqref{ham_uni}, \eqref{ham_uni1}, only  three frequencies $\omega=\omega_0 , \omega_0 \pm \Delta$ need to be taken into account. The expressions for power , cold current and hot current (up to terms of the order $\lambda^2$ read as follows.

1) The power 
\begin{widetext}
\begin{gather}
\mathcal{P}=-\frac{\Delta}{\sum_{m,i=H,C}P_mG^{i}(\omega_{0}+m\Delta)(1+e^{-\frac{\omega_{0}+m\Delta}{T_{i}}})} \nonumber\\
\left(P_1P_0\left(G^{C}(\omega_{0}+\Delta)G^{C}(\omega_{0})\left(e^{-\frac{\omega_{0}+\Delta}{T_{C}}}-e^{-\frac{\omega_{0}}{T_{C}}}\right)+G^{C}(\omega_{0}+\Delta)G^{H}(\omega_{0})\left(e^{-\frac{\omega_{0}+\Delta}{T_{C}}}-e^{-\frac{\omega_{0}}{T_{H}}}\right)\right.\right.+\nonumber\\
G^{H}(\omega_{0}+\Delta)G^{C}(\omega_{0})\left(e^{-\frac{\omega_{0}+\Delta}{T_{H}}}-e^{-\frac{\omega_{0}}{T_{C}}}\right)+G^{C}(\omega_{0}-\Delta)G^{C}(\omega_{0})\left(e^{-\frac{\omega_{0}}{T_{C}}}-e^{-\frac{\omega_{0}-\Delta}{T_{C}}}\right) \nonumber\\
+G^{H}(\omega_{0}-\Delta)G^{C}(\omega_{0})\left(e^{-\frac{\omega_{0}}{T_{C}}}-e^{-\frac{\omega_{0}-\Delta}{T_{H}}}\right)+G^{C}(\omega_{0}-\Delta)G^{H}(\omega_{0})\left(e^{-\frac{\omega_{0}}{T_{H}}}-e^{-\frac{\omega_{0}-\Delta}{T_{C}}}\right)+ \nonumber\\
\left.G^{H}(\omega_{0}+\Delta)G^{H}(\omega_{0})\left(e^{-\frac{\omega_{0}+\Delta}{T_{H}}}-e^{-\frac{\omega_{0}}{T_{H}}}\right)+G^{H}(\omega_{0}-\Delta)G^{H}(\omega_{0})\left(e^{-\frac{\omega_{0}}{T_{H}}}-e^{-\frac{\omega_{0}-\Delta}{T_{H}}}\right)\right)+ \nonumber\\
2P_1^{2}\left(G^{C}(\omega_{0}+\Delta)G^{C}(\omega_{0}-\Delta)\left(e^{-\frac{\omega_{0}+\Delta}{T_{C}}}-e^{-\frac{\omega_{0}-\Delta}{T_{C}}}\right)+G^{C}(\omega_{0}+\Delta)G^{H}(\omega_{0}-\Delta)\left(e^{-\frac{\omega_{0}+\Delta}{T_{C}}}-e^{-\frac{\omega_{0}-\Delta}{T_{H}}}\right)+\right.\nonumber\\
\left.\left.G^{H}(\omega_{0}+\Delta)G^{C}(\omega_{0}-\Delta)\left(e^{-\frac{\omega_{0}+\Delta}{T_{H}}}-e^{-\frac{\omega_{0}-\Delta}{T_{C}}}\right)+G^{H}(\omega_{0}+\Delta)G^{H}(\omega_{0}-\Delta)\left(e^{-\frac{\omega_{0}+\Delta}{T_{H}}}-e^{-\frac{\omega_{0}-\Delta}{T_{H}}}\right)\right)\right)
\end{gather}
\end{widetext}

2) The hot current
\begin{widetext}
\begin{gather}
J_{H}=
\frac{1}{\sum_{m,i=H,C}P_mG^{i}(\omega_{0}+m\Delta)(1+e^{-\frac{\omega_{0}+m\Delta}{T_{i}}})}\nonumber\\
\left( P_0^{2}\omega_{0}G^{H}(\omega_{0})G^{C}(\omega_{0})\left(e^{-\frac{\omega_{0}}{T_{H}}}-e^{-\frac{\omega_{0}}{T_{C}}}\right)+
P_1P_0\left(-\omega_{0}G^{C}(\omega_{0}+\Delta)G^{H}(\omega_{0})\left(e^{-\frac{\omega_{0}+\Delta}{T_{C}}}-e^{-\frac{\omega_{0}}{T_{H}}}\right)\right.\right.+\nonumber\\
(\omega_{0}+\Delta)G^{H}(\omega_{0}+\Delta)G^{C}(\omega_{0})\left(e^{-\frac{\omega_{0}+\Delta}{T_{H}}}-e^{-\frac{\omega_{0}}{T_{C}}}\right)
\nonumber\\
-(\omega_{0}-\Delta)G^{H}(\omega_{0}-\Delta)G^{C}(\omega_{0})\left(e^{-\frac{\omega_{0}}{T_{C}}}-e^{-\frac{\omega_{0}-\Delta}{T_{H}}}\right)
+\omega_{0}G^{C}(\omega_{0}-\Delta)G^{H}(\omega_{0})\left(e^{-\frac{\omega_{0}}{T_{H}}}-e^{-\frac{\omega_{0}-\Delta}{T_{C}}}\right)+\nonumber\\
\left.\Delta G^{H}(\omega_{0}+\Delta)G^{H}(\omega_{0})\left(e^{-\frac{\omega_{0}+\Delta}{T_{H}}}-e^{-\frac{\omega_{0}}{T_{H}}}\right)+
\Delta G^{H}(\omega_{0}-\Delta)G^{H}(\omega_{0})\left(e^{-\frac{\omega_{0}}{T_{H}}}-e^{-\frac{\omega_{0}-\Delta}{T_{H}}}\right)\right)+\nonumber\\
P_1^{2}\left(-(\omega_{0}-\Delta)G^{C}(\omega_{0}+\Delta)G^{H}(\omega_{0}-\Delta)\left(e^{-\frac{\omega_{0}+\Delta}{T_{C}}}-e^{-\frac{\omega_{0}-\Delta}{T_{H}}}\right)+(\omega_{0}+\Delta)G^{H}(\omega_{0}+\Delta)G^{C}(\omega_{0}-\Delta)\left(e^{-\frac{\omega_{0}+\Delta}{T_{H}}}-e^{-\frac{\omega_{0}-\Delta}{T_{C}}}\right)\right.\nonumber\\
+2\Delta G^{H}(\omega_{0}+\Delta)G^{H}(\omega_{0}-\Delta)\left(e^{-\frac{\omega_{0}+\Delta}{T_{H}}}-e^{-\frac{\omega_{0}-\Delta}{T_{H}}}\right)+\nonumber\\
\left.\left.(\omega_{0}-\Delta)G^{H}(\omega_{0}-\Delta)G^{C}(\omega_{0}-\Delta)\left(e^{-\frac{\omega_{0}-\Delta}{T_{H}}}-e^{-\frac{\omega_{0}-\Delta}{T_{C}}}\right)\right)\right)
\end{gather}
\end{widetext}
3) The cold current expression can be obtained from $J_H$ by interchanging  C and H.

\setcounter{equation}{0}
\renewcommand{\theequation}{B\arabic{equation}}
\subsection*{Appendix B. Master Equation}

To second order in the system-bath coupling, the  non-Markovian master-equation for the reduced system density matrix, $\rho_S(t)$ has the form \cite{KofmanPRL04}:
\begin{multline}
\label{gen-ME}
\dot{\rho_S}(t) = \\
-i\left[H_S,\rho_S(t)\right]+\\
\int_0^t d\tau \left\{
\Phi_T(t-\tau) \left[\tilde{S}(t,\tau)\rho_S(t),\sigma_x\right] +H.c.
\right\}
\end{multline}
Here  $\tilde{S}(t,\tau) = e^{-iH_S(t-\tau)}\sigma_xe^{iH_S(t-\tau)}$
 and the bath autocorrelation function is
$
\Phi_T(t) = \epsilon^2\langle B e^{-iH_Bt}Be^{iH_Bt} \rangle_B
$
,  $\epsilon$ being the coupling strength.

At equilibrium $\rho_S$ is diagonal in the energy basis of the TLS ($|e\rangle$, $|g\rangle$) and it remains so under the action of the diagonal $H_S$ (Eq. (3)). The rotation-wave approximation is not assumed here and hence 
\eqref{gen-ME} allows for arbitrarily fast modulations of the system. 
The corresponding rate equation are then given by Eqs. (4), with time-dependent rates $R_{g(e)}(t)$ given by Eq. (5).


%

\end{document}